# COMBUST: Gridded combustible mass estimates of the built environment in the conterminous United States (1975-2020)

Johannes H. Uhl[1,2,3], Maxwell C. Cook[2,4], Cibele Amaral[2,5], Stefan Leyk[1,4], Jennifer K. Balch[2,4,5], Alan Robock[6], Owen B. Toon[7]

1. University of Colorado Boulder, Institute of Behavioral Science, 483 UCB, Boulder, CO-80309, USA.
2. University of Colorado Boulder, Earth Lab, Cooperative Institute for Research in Environmental Sciences (CIRES), 216 UCB, Boulder, CO-80309, USA.
3. European Commission, Joint Research Centre (JRC), Via E. Fermi 2749, 21027 Ispra, Italy
4. University of Colorado Boulder, Department of Geography, 260 UCB, Boulder, CO-80309, USA.
5. University of Colorado Boulder, Environmental Data Science Innovation & Inclusion Lab (ESIIL), 216 UCB, Boulder, CO-80309, USA.
6. Department of Environmental Sciences, Rutgers University, 14 College Farm Rd, New Brunswick, NJ-08901, USA.
7. Department of Atmospheric and Oceanic Sciences, Laboratory for Atmospheric and Space Physics, University of Colorado Boulder, 311 UCB, Boulder, CO-80309, USA.

Corresponding author: Johannes H. Uhl (johannes.uhl@colorado.edu)

**Abstract:** The increasing occurrence of natural hazards such as wildfires and drought, along with urban expansion and land consumption, causes increasing levels of fire risk to populations and human settlements. Moreover, increasing geopolitical instability in many regions of the world requires evaluation of scenarios related to potential hazards caused by military operations. Quantitative knowledge on burnable fuels and their spatio-temporal distribution across landscapes is crucial for risk and potential damage assessments. While there is good understanding of the distributions of biomass fuels based on remote sensing observations, the combustible mass of the built environment has rarely been quantified in a spatially explicit manner. Therefore, we developed fine-grained estimates of urban fuels for the conterminous United States, estimating the combustible mass of building materials, building contents, and personal vehicles at 250 m spatial resolution. The resulting dataset is called COMBUST (Combustible mass of the built environment in the conterminous United States) and includes different backcasting scenarios from 1975 to 2020. COMBUST is based on the integration of a variety of geospatial data sources such as Earth-observation derived data, real estate data, statistical estimates and volunteered geographic information. COMBUST is accompanied by COMBUST PLUS, a set of consistently enumerated gridded datasets facilitating combustion exposure modelling of buildings and population. These datasets constitute a rich resource for ecological and social science applications, as well as for disaster risk management and planning-related decision making for U.S. settlements. COMBUST is available at https://doi.org/10.5281/zenodo.15611963.

## 1. Introduction

The increasing occurrence of natural hazards such as wildfires and drought, alongside expanding rural and urban development and land conversion, causes increasing levels of fire risk to populations and human settlements, particularly where they intermingle with vegetation, known as the wildland-urban interface (WUI) (Steward et al., 2007, Radeloff et al., 2018, Mietkiewicz et al., 2020, Cheng et al., 2024). Moreover, increasing geopolitical instability in many regions of the world requires reconsideration and risk evaluation of different scenarios of military operations and their potential impact on society. These developments catalyze the need for improved assessments of the exposure of people and infrastructure and thus quantitative, fine-grained information on the human habitat, i.e., the built environment. While there is a good understanding of human settlements in urban areas, in particular in data-rich countries, our quantitative knowledge on where people live and how people build is surprisingly sparse. These limitations impede the consistent measurement of long-term changes in the built environment over extended time periods and across large spatial extents. Specifically, building-level information relevant to natural hazards or potential outcomes of political conflict, such as building material, building volume, and the mass of built-up structures is scarce. Such data are crucial for risk and damage assessment, given the rapid growth of urban and rural (informal) settlements, increasingly dense and heterogeneous urban spaces, and the urgent need for building resilient and sustainable infrastructure.

Of particular interest are estimates of volume and material of built-up structures, which can be used to determine their fuel mass, or combustible mass (Frishcosy et al. 2021). While biomass fuel has been widely studied and quantified, using remotely sensed Earth observation data (Burgan et al., 1998, Reeves et al., 2009), the fuel of built-up structures is rarely quantified, often due to a lack of suitable data. Yet, in just the last five years the U.S. has seen more than $81 billion in losses and more than 60,000 structures destroyed by wildfires (St. Denis et al., 2023). For a seamless and holistic understanding of fuel distributions along the wildland-urban interface, such data are urgently needed. For example, such fuel estimates can be used to quantify the impacts of wildfires on the built environment, to estimate emissions and their impact on air quality (National Academy of Sciences, 2022), to model fire spread between adjacent built-up structures (Mahmoud & Chulahwat, 2018), and its interactions with vegetation-based fire spread. Hence, there is a pressing need to quantify the flammability of the built environment, and monitor these characteristics over time, as an important metric for integrated risk assessments and a determinant for sustainable and fire-resistant development (Smith et al., 2016, McWethy et al., 2019).

Moreover, information about building mass is crucial to assess the impact of development on the environment within urban and peri-urban ecosystems or close to natural habitats. Such information could be used for research on the impact of development on vegetation, hydrological systems, micro-climate as well as wildlife. Building mass is also critical for the estimation of expected building debris and rubble in the case of a natural disaster such as earthquakes or landslides (Ural et al., 2011, Du et al., 2013) to facilitate and support effective disaster response and risk management.

Combined aspects of flammability, resilience and mass are also crucial in efforts to estimate the potential impact in scenarios of war, armed conflict and missile strikes. In such scenarios it is important to have an understanding of possible damages, the material that might burn and the amount or mass of the built environment.

For example, studies of the environmental impacts of nuclear wars show that smoke and soot from burning cities can reach the stratosphere in sufficient quantities to produce devastating climate changes, threatening the bulk of humanity with agricultural disaster and starvation (Turco et al., 1983, 1990, Robock et al., 2007a,b, Stenke et al., 2013, Pausata et al., 2016, Coupe et al. 2019, Toon et al., 2019, Xia et al., 2022). These studies require estimates of fuel loads for urban materials at high spatial coverage. Since the emission factors vary with the type of material burning, the fuel load needs to differentiate a variety of materials including wood and lumber, paper, plastics and polymers, hydrocarbons, asphalt, and cloth. Previous studies, such as Turco et al. (1990), primarily found fuel loads from top-down estimates of the total amounts of material in use. Toon et al. (2007, 2008) summarised this approach by assigning 11 metric tons of fuel to each person in the developed world. The types of fuels were assumed constant and based on the top-down surveys of types of fuels. Fuel loads for a small number of cities were measured indicating that fuel loads were proportional to population density (Toon et al., 2007). However, building materials vary from place to place, the types of fuel in buildings has changed in time, in particular as the use of plastics has increased, and the amount of fuel per person may vary greatly with population density, wealth, and cultural trends. Therefore, detailed data of fuel load at fine spatial and temporal resolutions are needed to estimate the impacts of war-related disasters, in a more sophisticated manner at regional and local scales.

While fine-grained data on built-up areas can be obtained from Earth observation data (e.g., Pesaresi et al., 2024, Marconcini et al., 2020a), covering relatively long temporal windows (i.e., since the 1970s or 1980s; Gong et al., 2020, Marconcini et al., 2020b, Pesaresi et al., 2024), only recently, large-scale mapping efforts on the vertical component of the built environment have been initiated (Pesaresi et al, 2021, Esch et al., 2022, Che et al., 2024, Ma et al., 2024, Florio et al., 2025). While such data enable the geometric description of the built environment at fine grain, data on other properties such as building material or function are rare and do not capture temporal dynamics (Haberl et al., 2021, Frantz et al., 2023, Haberl et al., 2025). However, longitudinal urban data is key to fully understand urban dynamics (Papini et al., 2025), and knowledge of these properties and their changes over time is crucial for the quantification and evolution of mass and flammable material contained in buildings. Hence, alternative data sources need to be exploited, such as real-estate, cadastral or tax assessment data, often containing field survey or manually verified data. Such data are increasingly made available to the public, by governmental authorities (Uhl & Leyk, 2022) or via industry-fuelled data production efforts (Leyk et al., 2020) and offer unique opportunities for scientific studies of the built environment (Nolte et al., 2023).

Herein, we make use of different data sources, to quantify and evaluate the combustible mass of the built environment in the conterminous United States (CONUS) over time. Specifically, we integrate the "Material stock map of CONUS" dataset (MSMC; Frantz et al. 2023), the Zillow Transaction and Assessment Dataset (ZTRAX; Zillow 2020), Microsoft's USBuildingFootprints dataset (Microsoft 2021), historical building volume and population data from the Global Human Settlement Layer (GHSL; Pesaresi et al., 2024), as well as historical building densities from the HISDAC-US dataset (Leyk & Uhl, 2018), among other data sources, enumerated in grids of 250 m × 250 m, semidecadally from 1975 to 2020, and call this dataset COMBUST (**CO**mbustible **M**ass of the **B**uilt Environment in the **U**nited **ST**ates). COMBUST is publicly available and is accompanied by COMBUST Plus, a set of gridded surfaces for population and building exposure analysis to be used in conjunction with COMBUST. Herein, we describe the COMBUST dataset (Section 2), the source data and methods to produce COMBUST (Section 3), showcase the data and high-level findings (Section 4). Section 5 concludes with a discussion of limitations of COMBUST.

## 2. Dataset description

COMBUST consists of a large set of almost 200 gridded datasets in GeoTIFF format, referenced in Albers Equal Area Conic projection for the conterminous US (EPSG:5070), enumerated in grid cells of 250m x 250m, aligned to the grid of the Historical Settlement Data Compilation for the U.S. (HISDAC-US, Leyk and Uhl, 2018). All mass estimates are totals reported in tons per grid cell. The suffixes "low", "mean", and "high" denominate the three scenarios implemented in the MSMC data. COMBUST is organized in 9 ZIP archives, one for each of the nine themes (see below).

### 2.1. Components / themes of the COMBUST dataset:

1. Combustible mass by component: building contents, building material, personal vehicles, gas stations, refineries. (2020)
2. COMBUST-PLUS: Thematic layers of the built environment and biomass for exposure and interaction modelling (2010-2020)
3. Combustible mass of buildings and their content by material (2020)
4. Combustible mass of buildings and their content by material type (2020)

5. Historical estimates of combustible mass of personal vehicles (1975-2020), by material (1995-2020)
6. Historical estimates of combustible mass of buildings - 1975-2020 (5-yr intervals), model 1 (using HISDAC-US building indoor area change rates)
7. Historical estimates of combustible mass of buildings - 1975-2020 (5-yr intervals), model 2 (using HISDAC-US historical building density change rates)
8. Historical estimates of combustible mass of buildings - 1975-2020 (5-yr intervals), model 3 (using GHSL building volume change rates)
9. Mass of non-combustible materials for rubble and debris estimation (2020)

## 2.2. Detailed list of COMBUST data records

| Theme | Layer description | File names |
|---|---|---|
| 1 | Total comb. mass of the built environment incl. cars | combust_cm_total_scenario_<low,mean,high>_2020.tif |
| 1 | Comb. mass of building material | combust_cm_buildingmaterial_all_scenario_<low,mean,high>_2020.tif |
| 1 | Comb. mass of building contents | combust_cm_buildingcontent_total_2020.tif |
| 1 | Comb. mass of fuel in gas stations | combust_cm_gasstations_2020.tif |
| 1 | Comb. mass of fuel in refineries | combust_cm_refineries_2020.tif |
| 1 | Comb. mass of cars | combust_cm_car_total_t_2020.tif |
| 2 | Number of buildings (Source: Microsoft USBuildingFootprints) | combust_plus_num_buildings_2020.tif |
| 2 | Total built-up area (Source: Microsoft USBuildingFootprints) | combust_plus_builtup_area_2020.tif |
| 2 | Number of residential units (Source: HISDAC-US V2) | combust_plus_num_units_2020.tif |
| 2 | Share of residential buildings (Source: HISDAC-US V2) | combust_plus_residential_building_share_2020.tif |
| 2 | Average cadastral parcel size (Source: ZTRAX) | combust_plus_average_lotsize_2020.tif |
| 2 | Average building construction year (Source: ZTRAX) | combust_plus_average_constr_year_2020.tif |
| 2 | Earliest building construction year (Source: ZTRAX) | combust_plus_earliest_constr_year_2020.tif |
| 2 | Local property ownership rate (Source: ZTRAX) | combust_plus_local_ownership_rate_2020.tif |
| 2 | Resident population (Source: GHSL R2023A, GHS-POP) | combust_plus_resident_population_<1975-2020>.tif |
| 2 | Above-ground combustible biomass (Source: Spawn et al. 2020) | combust_plus_combustible_biomass_aboveground_2010.tif |
| 2 | Below-ground combustible biomass (Source: Spawn et al. 2020) | combust_plus_combustible_biomass_belowground_2010.tif |
| 2 | Total combustible biomass (Source: Spawn et al. 2020) | combust_plus_combustible_biomass_total_2010.tif |
| 3 | Mass of combustible building contents: cloth component | combust_cm_buildingcontent_cloth.tif |
| 3 | Mass of combustible building contents: paper component | combust_cm_buildingcontent_paper.tif |
| 3 | Mass of combustible building contents: plastic component | combust_cm_buildingcontent_plastic.tif |
| 3 | Mass of combustible building contents: wood component | combust_cm_buildingcontent_wood.tif |
| 3 | Mass of combustible building materials: petrochemical-based materials | combust_cm_buildingmaterial_all_other_petrochemical_based_materials_scenario_<low,mean,high>.tif |
| 3 | Mass of combustible building materials: bitumen | combust_cm_buildingmaterial_bitumen_scenario_<low,mean,high>.tif |
| 3 | Mass of combustible building materials: other biomass-based materials | combust_cm_buildingmaterial_other_biomass_based_materials_scenario_<low,mean,high>.tif |
| 3 | Mass of combustible building materials: timber | combust_cm_buildingmaterial_timber_scenario_<low,mean,high>.tif |
| 4 | Comb. mass of buildings (comm./indust.) | combust_cm_building_commercial_industrial_scenario_<low,mean,high>_2020.tif |
| 4 | Comb. mass of buildings (comm. - inner city.) | combust_cm_building_commercial_innercity_scenario_<low,mean,high>_2020.tif |
| 4 | Comb. mass of buildings (highrise) | combust_cm_building_highrise_scenario_<low,mean,high>_2020.tif |
| 4 | Comb. mass of buildings (lightweight) | combust_cm_building_lightweight_scenario_<low,mean,high>_2020.tif |
| 4 | Comb. mass of buildings (multifamily) | combust_cm_building_multifamily_scenario_<low,mean,high>_2020.tif |
| 4 | Comb. mass of buildings (single family) | combust_cm_building_singlefamily_scenario_<low,mean,high>_2020.tif |
| 4 | Comb. mass of buildings (skyscraper) | combust_cm_building_skyscraper_scenario_<low,mean,high>_2020.tif |
| 5 | Historical estimates of total comb. mass of personal cars of residents | combust_cm_car_total_t_<Year>.tif |
| | Historical estimates of comb. mass of personal cars of residents, plastic | combust_cm_car_plastic_t_<Year>.tif |
| | Historical estimates of comb. mass of personal cars of residents, rubber | combust_cm_car_rubber_t_<Year>.tif |
| | Historical estimates of comb. mass of personal cars of residents, fluids | combust_cm_car_fluidlubricants_t_<Year>.tif |
| 6 | Historical estimates of comb. mass of building since 1975, model 1, non-backcastable combustible mass | combust_cm_building_all_scenario_<low,mean,high>_backcasted_mod1_hisdacus_bui_<Year>.tif |
| 7 | Historical estimates of comb. mass of building since 1975, model 2, non-backcastable combustible mass | combust_cm_building_all_scenario_<low,mean,high>_backcasted_mod2_hisdacus_bupl_<Year>.tif |
| 8 | Historical estimates of comb. mass of building since 1975, model 3, non-backcastable combustible mass | combust_cm_building_all_scenario_<low,mean,high>_backcasted_mod3_ghsl_<Year>.tif |
| 9 | Non-combustible mass of cars, historical estimates | combust_noncombust_car_non_combmass_t_<Year>.tif |
| | Non-combustible mass of building materials | combust_noncombust_<low,mean,high>_noncombust_total_mass_ext_t.tif |

## 3. Methodology

### 3.1. Combustible mass of buildings

Combustible mass of the building stock ($CM_{Building\ stock}$) is represented as

$$CM_{Built\ environment} = CM_{Building\ material} + CM_{Building\ content} + CM_{Gas\ stations} + CM_{Refineries}$$

The combustible mass of **building material** has been derived from data on building volume and building mass, available for the conterminous United States at 10m spatial resolution for the year 2018, called "Material stock map of CONUS" (Frantz et al. 2023). These data, henceforth shortened to MSMC, were aggregated to the 250m target grid, and non-flammable materials were excluded. The mass of the following materials was taken into account: timber and other biomass-based materials, bitumen and other petrochemical-based materials use for the construction of buildings. Frantz et al. Used (a) Microsoft building footprint data for the US to delineate buildings, (b) they estimated building height and building type from Sentinel-1 and Sentinel-2 Earth observation data. Based on these data, they estimated building volume per 10m grid cell, and used estimates of the material composition of each building type, to estimate the mass of buildings (enumerated in 10m grid cells). Frantz et al. estimate a lower and an upper bound of building material mass, and report also the average of both. These three estimates are reflected in the "low", "mean", and "high" variants of the COMBUST surfaces. COMBUST reports the mass of all combustible materials, per material type, per building type, and overall. The detailed process is described in Appendix 1.

### 3.2. Combustible mass of building contents

The combustible mass of **building contents** has been derived using a method proposed by Frishcosy et al. (2021). This method infers the mass of flammable building contents based on building indoor area and building function. We used property-level information from the Zillow Transaction and Assessment Dataset (ZTRAX), commonly used for environmental research (Nolte et al. 2024), for the year 2020, providing building indoor area and function for large parts of the properties in the US. Part of the authors had access to the ZTRAX dataset in the scope of a data share agreement. By applying the method from Frishcosy et al., an estimate of the combustible mass of building content was made, which was aggregated to the 250 m target grid (see Appendix 1 for details). As the Frantz et al. Building mass data is referenced to 2018, and the ZTRAX data underlying the building content mass estimates is referenced to 2020, our estimates reflect the state of the building stock between 2018 and 2020, in average 2019. For grid cells where no ZTRAX data was available (affecting roughly 3% of CONUS counties), the combustible mass of building contents was imputed using a linear regression model estimating $CM_{Building\ content}$ as a function of $CM_{Building\ material}$, i.e., $CM_{Building\ content} = a + b*CM_{Building\ material}$. The detailed process is described in Appendix 2.

### 3.3. Combustible mass of gas stations and refineries

The **combustible mass of gas stations** was estimated separately. From OpenStreetMap, the geolocations of almost 100,000 gas stations in the CONUS were obtained, from an OSM database version as of Sept 2024. Based on an estimate of fuel stored in gas stations in average, the combustible mass of fuel reserves in gas stations was estimated, and added to the grid cells in which the gas stations are located. Similarly, the **combustible mass of refineries** was estimated based on geolocation data from the Homeland Infrastructure Foundation-Level Database (HIFLD 2025), published in 2021, using an estimated stock of fuels stored at refinery locations, in average. Likewise, the estimated combustible mass was added to the grid cells in which the refineries are located.

### 3.4. Combustible mass of personal vehicles

We used statistics on the number of **personal vehicles** per capita, over time (Davis and Boundy 2021) and on the average weight and material composition of personal vehicles over time. From these materials, we selected combustible materials, and used gridded data on residential population distribution from the Global Human Settlement Layer (GHSL), i.e., the GHS-POP R2023A dataset to estimate the distributions of combustible car material per 250m grid cell, by estimating the number of personal vehicles per grid cell (based on the average number of personal vehicles per capita and total resident population per cell) Based on the estimated number of personal vehicles per grid cell, we assigned the corresponding weight of combustible car materials to each grid cell.

### 3.5. Backcasting of building-related combustible mass to 1975

We implemented three models for backcasting of combustible mass from 1975 to 2020 in 5-year intervals:

- **Model 1:** Uses the historical change rates of a conflated version of HISDAC US V1 and V2 Built-up intensity (BUI) indoor area estimates (Leyk & Uhl 2018) to backcast the 2020 gridded combustible mass data.
- **Model 2:** Uses the historical change rates of a conflated version of HISDAC US V1 and V2 built-up property locations (BUPL), a measure of building density, to backcast the 2020 gridded combustible mass data (Uhl et al., 2021).
- **Model 3:** Uses the historical change rates gridded building volme estimates from the Global Human Settlement Layer (GHSL; GHS-BUILT-V R2023A) to backcast the 2020 gridded combustible mass data (Pesaresi et al., 2024).

From these datasets, the change rates with respect to the year 2020 were extracted, and applied to the 2020 combustible mass estimates. While the GHSL-based backcasting results are spatially nearly exhaustive, the HISDAC-US based backcasting results suffer from lower levels of coverage, as the historical information (derived from building construction year information from ZTRAX underlying the HISDAC-US data) is not available in some states, including parts of New Mexico, the Dakotas, Wisconsin, and Louisiana.

### 3.6. Non-combustible mass estimates of buildings and cars

The materials of buildings and cars that are considered non-combustible were aggregated separately to estimate the total non-combustible mass of buildings and personal vehicles per 250m grid cell, facilitating rubble and debris estimations.

### 3.7. COMBUST-Plus – accompanying population and building exposure variables, and combustible biomass estimates

Data on biomass carbon density, produced by Spawn et al. (2020) for the year 2010 was used to estimate the total combustible biomass per 250m grid cell. These data were resampled from per-hectare density estimates. The combustible biomass was estimated to be 2x the carbon mass reported by Spawn et al. (2020), and this process was done separately for above-ground and below-ground biomass estimates.

Further variables in COMBUST-Plus include

- Building density and built-up area per 250m grid cell, as reported in Microsoft´s USBuildingFootprint dataset, and the average cadastral parcel size per grid cell, as reported in ZTRAX, as measures of built-up intensity.

- Number and share of residential building units, as reported in HISDAC-US V2 (Ahn et al., 2024), to quantify the exposure of residential properties.

- The total population per 250m grid cell, as reported in GHSL, and the share of locally owned properties, as derived from ZTRAX. The latter allows to estimate potentially seasonally uninhabited residential units, for example in settlements dominated by vacation homes.

- Average and earliest construction year per 250m grid cell, as derived from HISDAC-US, measuring the age of the built stock that can be possibly linked to building characteristics related to building resilience and value.

These accompanying variables allow for quantification of exposed population, structures, and potentially to estimate building resilience and / or damage.

The complete data processing workflow is shown in Fig. 1.

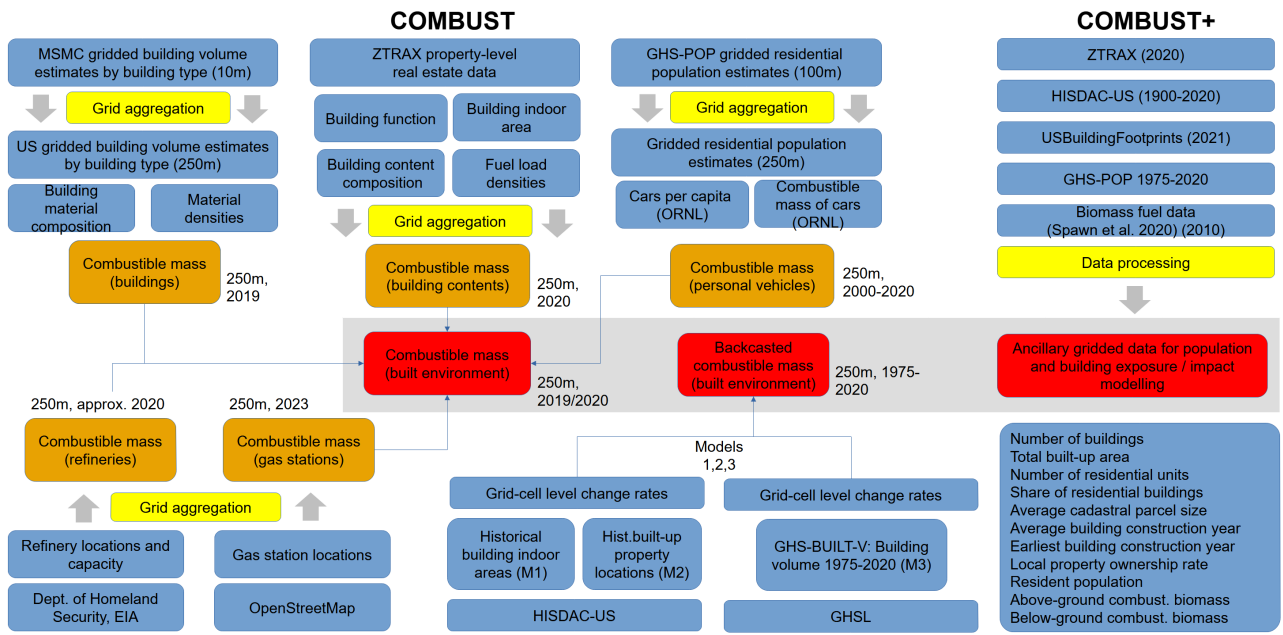

*Figure 1. COMBUST data processing workflow.*

## 4. Results

### 4.1. Spatial distributions of combustible mass across the conterminous U.S.

Exemplary gridded surface measuring combustible mass are shown in Figures 2, 3, and 4. Figure 2 shows the total combustible mass (including building materials, contents, gas stations, refineries, and personal vehicles, for the reference year 2020.

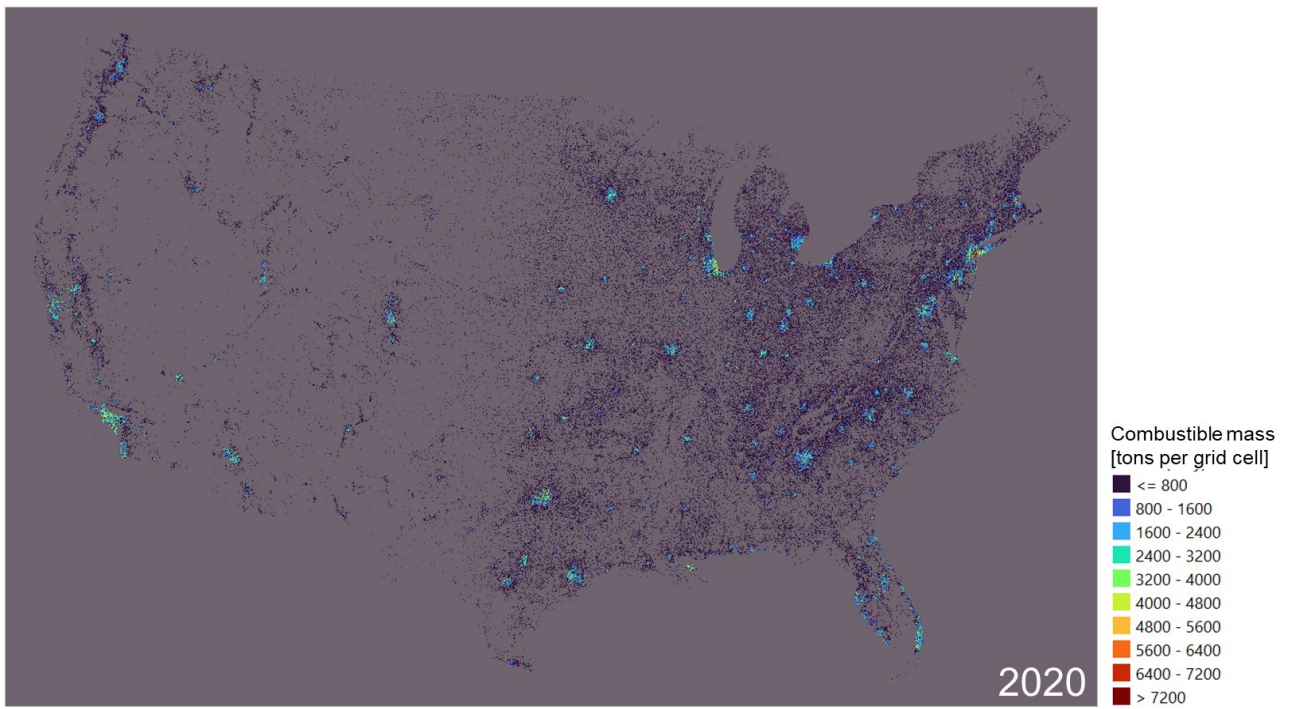

*Figure 2. COMBUST 250m gridded surface measuring combustible mass of buildings and their content in 2020.*

Figure 3 shows the total combustible mass (excluding personal vehicles) in 1975 and 2020, using Model 2 (i.e., HISDAC-US BUPL change rates) for backcasting. Figure 3 shows the distributions of combustible mass of building contents and personal vehicles for 2020, shown for the Greater Denver area.

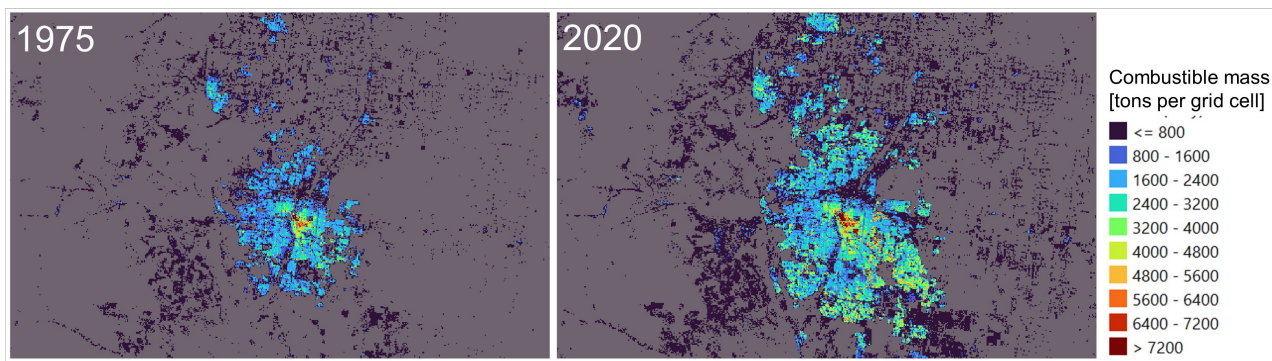

*Figure 3. COMBUST 250 m gridded surfaces measuring combustible mass of buildings and their content in 1975 and in 2020, shown for the Greater Denver area, Colorado.*

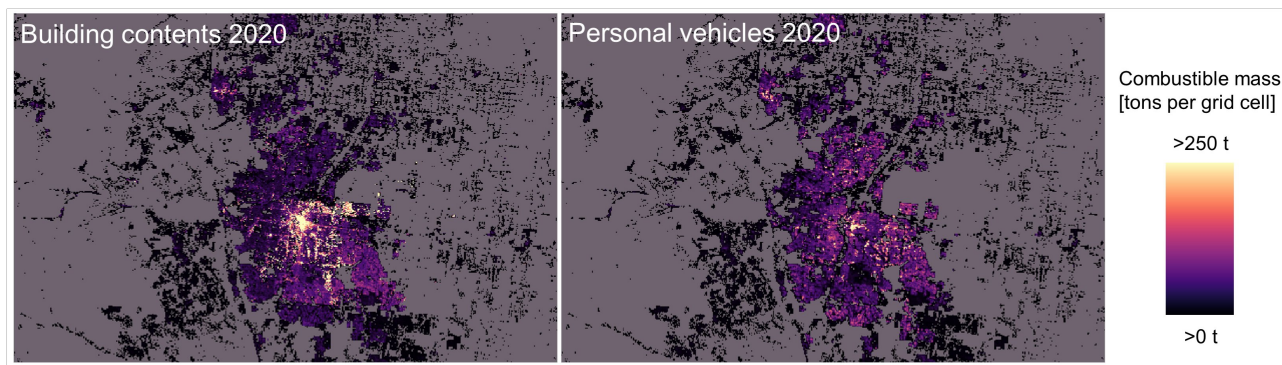

*Figure 4. COMBUST 250 m gridded surfaces measuring combustible mass of building contents (left) andof personal vehicles (right), in 2020, shown for Greater Denver area.*

**4.2. Dissecting combustible mass in 2020**

Frantz et al. estimate the total weight of the CONUS built environment to 127 Gt, out of which 62 Gt are the mass of building materials. We estimate a total mass of 64 Gt for building materials, personal vehicles and building contents. Out of these 64Gt, 55Gt are non-combustible, and only 7.8 Gt, corresponding to 13.7% are combustible. These 7.8 Gt represent the optimistic (upper bound) scenario from Frantz et al., while the average scenario results in a total of 6.9 Gt total combustible mass, corresponding to a total of 20.8 t of combustible mass per capita (CM/capita) in 2020. Out of these estimated 20.8 t/capita of combustible mass, around 18.2 t/capita are building material, 2.6 t/capita are estimated to be building contents, and 0.38 t/capita are combustible materials from personal vehicles (Table 1).

The large majority of the combustible mass of building materials are single-family residential homes (5.3 Gt out of 6.0 Gt in the "mean" scenario), corresponding to 16 t/capita, and the combustible mass of commercial and industrial buildings is estimated to 0.9 Gt (2.8 t/capita). Across the whole CONUS population in 2020, the combustible mass of high rise buildings and skyscrapers only makes up around 100kg/capita Table 1). For comparison, the mass of combustible materials contained in personal vehicles in 2020 is estimated almost four times higher (i.e., 387kg/capita; Table 2).

*Table 1. Total combustible (and non-combustible) mass and mass/capita by component.*

| Component | Mass (Gt) | | | Mass per capita (t/capita) | | |
|---|---|---|---|---|---|---|
| | low | mean | high | low | mean | high |
| Total mass (building materials, contents, personal vehicles) | 39.267 | 54.669 | 64.143 | 117.359 | 164.949 | 193.531 |
| Total non-combustible mass (building materials) | 32.687 | 46.490 | 55.014 | 98.617 | 140.262 | 165.981 |
| **Total combustible mass** | 5.289 | 6.911 | 7.846 | 15.957 | 20.850 | 23.671 |
| Comb. mass building material 2020 | 4.415 | 6.037 | 6.972 | 13.322 | 18.215 | 21.036 |
| Comb. mass building content 2020 | 0.874 | 0.874 | 0.874 | 2.635 | 2.635 | 2.635 |
| Comb. mass personal vehicles 2020 | 0.127 | 0.127 | 0.127 | 0.384 | 0.384 | 0.384 |
| Comb. mass gas stations 2020 | 0.010 | 0.010 | 0.010 | 0.031 | 0.031 | 0.031 |
| Comb. mass refineries 2020 | 0.001 | 0.001 | 0.001 | 0.003 | 0.003 | 0.003 |
| Comb. mass building_singlefamily 2020 | 4.167 | 5.304 | 6.041 | 12.572 | 16.002 | 18.227 |
| Comb. mass building_commercial_industrial 2020 | 0.782 | 0.921 | 0.996 | 2.360 | 2.780 | 3.004 |
| Comb. mass building_lightweight 2020 | 0.587 | 0.734 | 0.832 | 1.770 | 2.216 | 2.511 |
| Comb. mass building_multifamily 2020 | 0.417 | 0.561 | 0.600 | 1.259 | 1.694 | 1.812 |
| Comb. mass building_commercial_innercity 2020 | 0.086 | 0.111 | 0.111 | 0.258 | 0.334 | 0.334 |
| Comb. mass building_highrise 2020 | 0.021 | 0.026 | 0.026 | 0.064 | 0.078 | 0.080 |
| Comb. mass building_skyscraper 2020 | 0.013 | 0.015 | 0.015 | 0.040 | 0.046 | 0.046 |

*Table 2. Combustible mass overview of personal vehicles over time*

| | Mass (Mt) | | t/capita | | |
|---|---|---|---|---|---|
| Year | combustible | non-combustible | combustible | non-combustible | percent combustible |
| 1975 | 30.891 | 182.841 | 0.149 | 0.884 | 14% |
| 1990 | 53.235 | 258.267 | 0.219 | 1.065 | 17% |
| 2000 | 74.319 | 317.260 | 0.269 | 1.148 | 19% |
| 2020 | 127.375 | 368.045 | 0.387 | 1.119 | 26% |

## 4.3. Temporal trends of combustible mass

The combustible mass of personal vehicles (0.384 t/capita) in 2020 corresponds to only 1.6% of the total combustible mass. Since 1975, this value has almost tripled (from 0.149 t/capita to 0.384 t/capita), and the percent combustible mass of personal vehicles las increased from 14% in 1975 to 26% in 2020 (Table 2). The combustible mass per capita attributed to gas in gas stations and in refineries is even 1 and 2 orders of magnitude lower, respectively (Table 1).

Looking at the backcasting models, we observe increases of total combustible mass of the building stock roughly ranging between 40% and 50% during the period from 1975 to 2020, from 4 Gt in 1975 to 8 Gt in 2020 (Fig. 5a). Importantly, this increase was linear before 2000, while change rates decreased after 2000. Looking at combustible mass per capita over time (Fig.5b) across all CONUS, we see an increase in CM/capita from 1975 to approximately 2005, and a slight decrease since then. Our backcasting models assume growth of the built stock over time, disregarding potential shrinkage – hence, the decreasing CM/capita is attributed to growth rates higher for population than for the built stock.

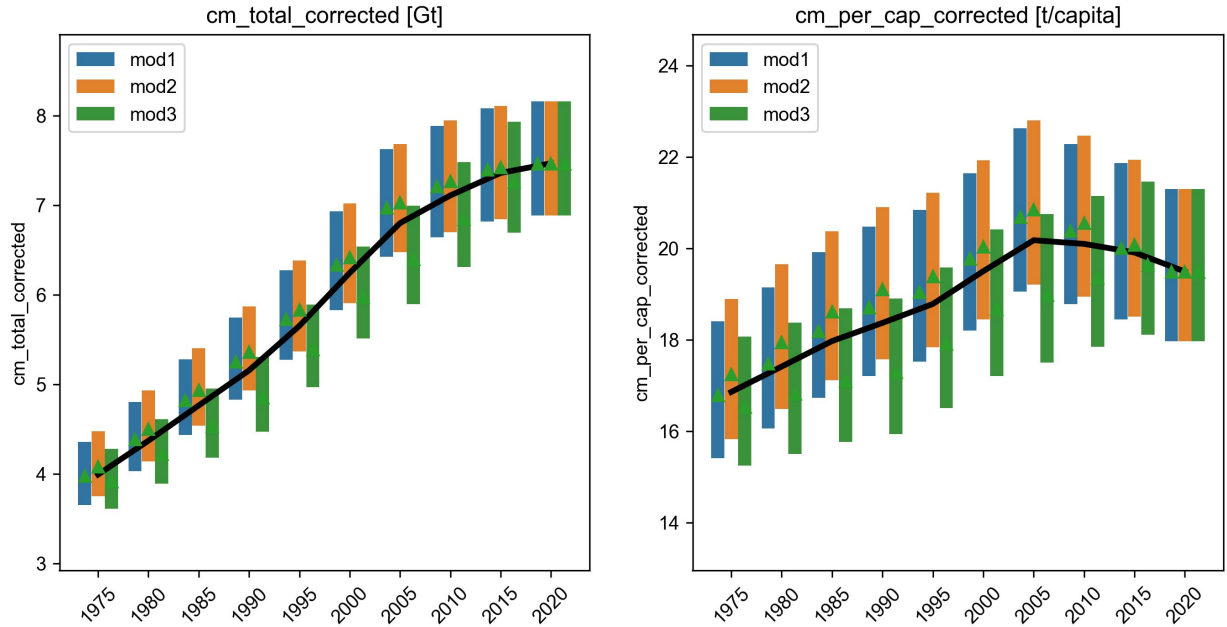

*Figure 5. Distributions of backcasted combustible mass estimates of building materials, contents, and personal vehicles in CONUS from 1975 to 2020: Boxplots (IQRs only) per backcasting model aggregated across total building mass scenarios from Frantz et al., for (a) total combustible mass, and (b) for combustible mass per capita.*

The estimates of total building mass provided by Frantz et al., on which the COMBUST estimates are based, are provided in three scenarios (low, mean, high), and this spread propagates also into the backcasted CM estimates (Figure 6).

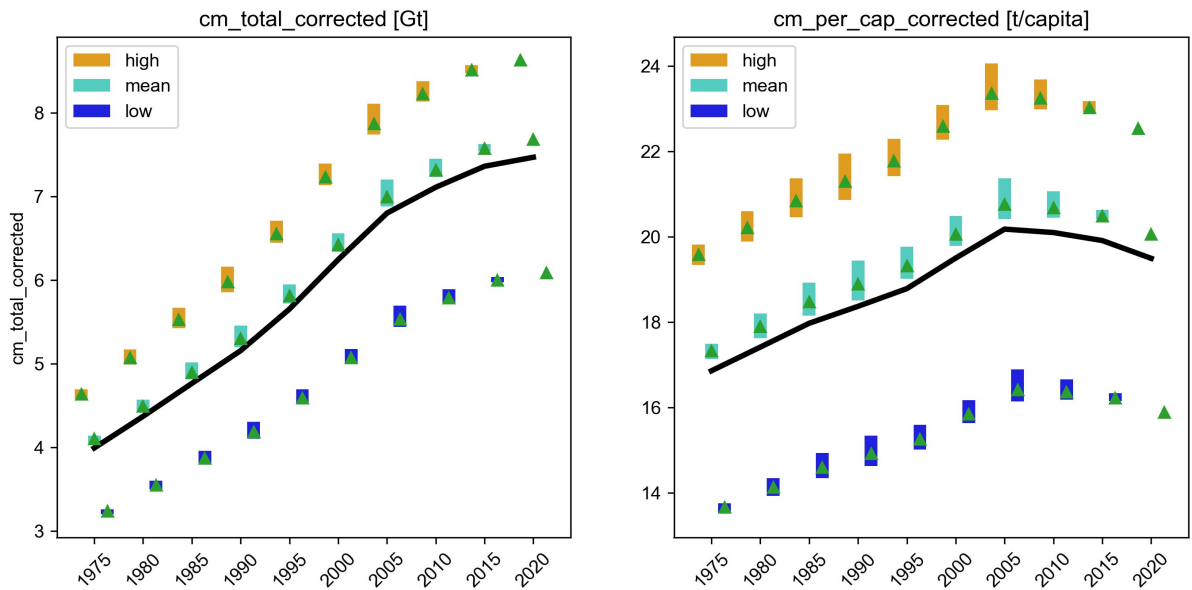

*Figure 6. Distributions of backcasted combustible mass estimates of building materials, contents, and personal vehicles in CONUS from 1975 to 2020: Boxplots (IQRs only) per total building mass scenario, aggregated across backcasting models, for (a) total combustible mass, and (b) for combustible mass per capita.*

We assume that the decreasing trend of CM/capita since 2005 is mostly attributed to higher population growth rates. Urbanization and rural outmigration cause urban populations to grow faster, and we assume this trend is dominated by urban strata: When disentangling the dynamics of CM/capita into metropolitan and non-metropolitan counties (according to the US Census´ core-based statistical area classification), as shown in Table 3: CM/capita in the "mean" scenario has grown from 24 t/capita to 34 t/capita in non-metro counties (a relative change of almost 100%), while the increase in metro counties from 18.7 t/capita in 1975 to 21.7 t/capita in 2020 corresponds to only 16% - while population in metro counties has increased by 65%, as compared to 27% in non-metro counties (Source: GHS-POP).

*Table 3. CM and CM/capita over time and in metro vs non-metro counties.*

| Year | Scenario | Metropolitan countries | | Non-metro counties | |
|------|----------|---------|----------|---------|----------|
|      |          | CM [Gt] | t/capita | CM [Gt] | t/capita |
| 1975 | Low  | 2.559 | 14.896 | 0.665 | 19.012 |
| 1975 | Mean | 3.221 | 18.746 | 0.839 | 23.985 |
| 1975 | High | 3.608 | 20.999 | 0.958 | 27.361 |
| 2020 | Low  | 4.906 | 17.249 | 1.211 | 27.220 |
| 2020 | Mean | 6.183 | 21.738 | 1.534 | 34.480 |
| 2020 | High | 6.923 | 24.340 | 1.748 | 39.302 |

This becomes evident further when looking across the rural-urban continuum, as modelled by USDA´s rural-urban continuum codes (RUCCs, Butler, 1990) (Figure 7): We find that combustible mass has increased across all rural-urban strata, and combustible mass is concentrated in urban strata. However, CM/capita is higher in rural counties than in urban counties, and CM/capita has largely been steadily increasing in most strata except in the two most urban strata, where we see the decrease in CM/capita after 2000, driving the signal observed in the overall trends across CONUS (cf. Figures 3,4). Steep increases of CM/capita in rural areas are likely a superposed effect of rural population decline alongside rural development by construction of new buildings.

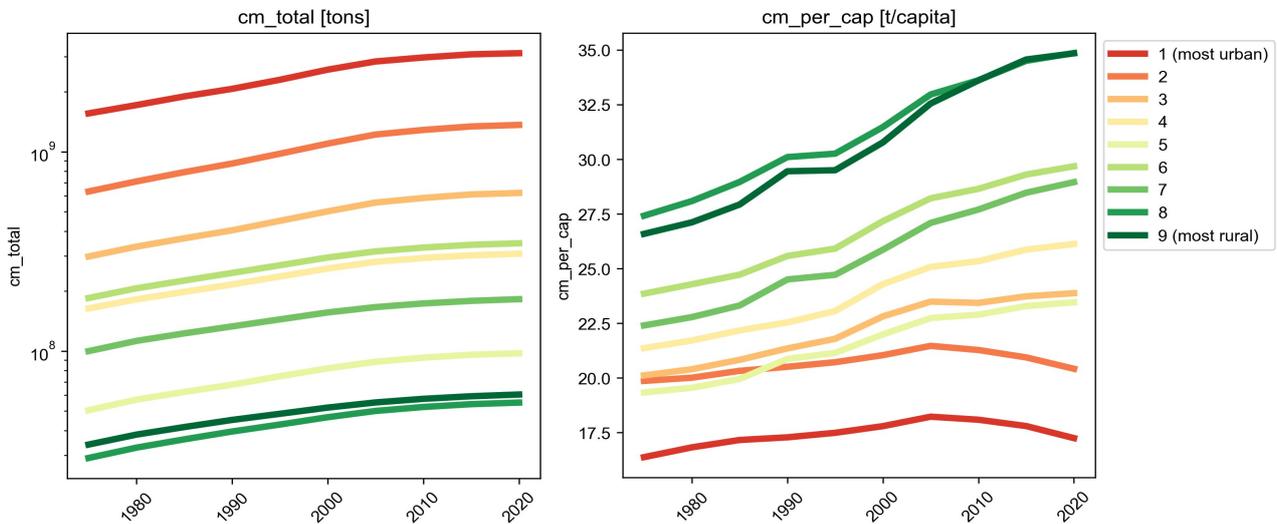

*Figure 7. Trends of combustible mass and CM/capita over time and across the rural-urban continuum, shown are totals for counties of each rural-urban continuum code*

Finally, we assess the results of the three backcasting models and the total building mass scenarios from Frantz et al. across rural-urban strata (Appendix Figures A3.1-A3.4), showing the same trends as observed in Figure 7, but also indicating that the CM/capita estimates in rural strata are affected by higher levels of uncertainty, as the spread between different models and scenarios is higher than in urban strata.

# 5. Limitations and conclusions

COMBUST represents a unique data source to quantify the combustible mass of the built environment, allowing for measuring building-related emissions in case of wildfires or other fire- or combustion-related hazards. The dataset is based on a multi-source data integration framework, conflating information from various remote-sensing based and other geospatial datasets.

While the results obtained from COMBUST are highly plausible, some limitations remain: COMBUST inherits uncertainty directly from the source data used from Frantz et al. (2023). Specifically, these data are affected by (a) uncertainty of thematic estimates of building function (e.g., residential, industrial, etc.), (b) uncertainty in the quantitative estimates of building mass and volume, and, potentially, (c) by positional uncertainty due to the quality of georeferencing of underlying remote sensing data. While the latter component has a minor effect due to the aggregation to 250m grid cells, uncertainty in building function estimates affects the model-based inference of building material composition. According to Frantz et al. (2023), building types are predicted with an overall accuracy of 79.92% (Frantz et al. 2023, Supplementary Table 7). Building mass in Frantz et al. Is derived from estimated building volume and material densities. Hence, uncertainty in building volume estimation (defined as building footprint area x building height) affects the quality of mass estimates. Building footprint area estimates were derived from Microsoft´s USBuildingFootprint data (Microsoft, 2021), estimated to be highly accurate (Microsoft, 2021). According to Frantz et al. (2023), their building height estimation based on Sentinel-1 and Sentinel-2 data achieves an average MAE of 2.99m (internal validation) and 3.21m (external validation). Based on building type and volume, and building-type specific material compositions and their densities, the total mass is estimated. This overall mass across the United States is estimated as 390 t/capita, which is in line with other studies, ranging from 300 t/capita to 430 t/capita (Supplementary Figure 9 of Frantz et al. 2023). While these reported accuracy metrics seem promising, and plausibility of these mass estimates is high, Frantz et al. (2023) provide upper and lower bounds, and their mean estimate for each grid cell. COMBUST is available for these three scenarios, allowing for evaluating the sensitivity of combustible mass estimates (Figure 6).

Additional uncertainty in COMBUST is introduced by the estimation of combustible mass of building contents using ZTRAX building indoor area and function, following a method proposed by Frishcosy et al. (2021). This method estimates the material composition of building contents based on building function, and estimates the total combustible mass using fuel load densities and building indoor area. ZTRAX covers the CONUS built environment well, and areas that are built-up but not covered by construction year information from ZTRAX correspond to approximately 2.7% of the CONUS land mass (Uhl et al., 2021). For the grid cells covered by the MSMC data, but not by ZTRAX, we imputed the combustible mass of building contents, introducing additional uncertainty. ZTRAX (or the derived HISDAC-US dataset) agrees well with Microsoft's USBuildingFootprints data along the rural-urban gradient (i.e., an F-score of 0.78 within census places in most rural counties, and >0.9 in most urban counties, Uhl et al., 2021), indicating high level of coherence between real-estate data and building footprint data derived from very high resolution satellite imagery.

For the estimation of historical building counts and building-level emissions from COMBUST, we used building densities derived from HISDAC-US historical building density data (i.e., the BUPL dataset). The accuracy of BUPL time series has been estimated to range above 0.8 (as measured by the F-score) between the years 1975 and 2000 in both rural and urban areas (see Uhl et al., 2021). The backcasting model 3 relies on historical building volume estimates from GHSL. While the vertical component of GHSL building volume estimates is difficult to validate (due to lack of multi-temporal 3D building data, the horizontal component (i.e., built-up surface) measured in GHSL has shown to be the most accurate as compared to other global, longitudinal settlement datasets (Pesaresi et al., 2024).

Moreover, the thematic accuracy of COMBUST for discriminating built-up and not built-up grid cells is high, and corresponds to the built-up domain of Microsoft´s USBuildingFootprint dataset (as these data were used as built-up mask in the underlying MSMC data (Frantz et al., 2023). A previous comparison of Microsoft data with authoritative building footprint data for a range of U.S. counties showed an F-score of 0.90 in low-density, and 0.97 in high-density counties (see Appendix F of Uhl et al., 2021), and these numbers are also applicable to the coverage of the built-up domain of COMBUST in 2020.

The total combustible mass per capita in CONUS, as reported in COMBUST, ranges between 15 and 23 t/capita, which is in the same order of magnitude as previous, top-down estimates (i.e., 11t/capita as reported by Toon et al. 2007, 2008). However, our estimates are slightly higher. This could be due to different data availability at the time, the approach used, and the inclusion of building contents in the COMBUST model. Furthermore, our estimates exhibit strong variations across the rural-urban gradient, which are in line with intra-urban variations found by Frishcosy et al. (2021).

Importantly, the backcasted COMBUST data do not take into account buildings that existed in the past but not in 2020 – building stock shrinkage and building stock renewal is not reflected in COMBUST. Concluding, COMBUST provides a valuable and plausible baseline dataset for assessing combustible mass in the context of wildfires and other hazards across spatio-temporal trajectories in the conterminous United States.

**Acknowledgments**

This work has been supported by the Open Philantropy Project (OPP). Moreover, we acknowledge access to the Zillow Transaction and Assessment Dataset (ZTRAX) through a data use agreement between the University of Colorado Boulder and Zillow Group Inc. More information on accessing the data can be found at http://zillow.com/ztrax. The results and opinions are those of the authors and do not reflect the position of Zillow Group. Support by Zillow Group Inc. is acknowledged.

**Data and code availability**

The COMBUST data are available at https://doi.org/10.5281/zenodo.15611963.

**Appendices**

**Appendix 1. Modelling combustible mass of building materials**

The MSMC data produced by Frantz et al. (2023) is available as 10-m raster data reporting the volume and mass of buildings, per building type. We first resampled (i.e., aggregated) the 10m volume estimates to the 250-m COMBUST grid. COMBUST is provided at 250m resolution as many ancillary data, such as HISDAC-US data, are available at that resolution. Based on accompanying information on material intensity factors ( in kg / m³; Baumgart et al., 2022), the cell-level total mass per building type was obtained from the total volume and disaggregated into individual material contributions, and re-aggregated into combustible mass per cell and building type, and non-combustible mass per cell and building type (Table A1.1).

*Table A1.1. Building types used in COMBUST. Source: Baumgart et al., 2022.*

| Building type (COMBUST) | Category (Baumgart et al.) | Description (from Baumgart et al. 2022) |
|---|---|---|
| singlefamily | RES-LR | (Semi-) Detached residential structure and attached small-scale residential or (rarely) commercial structure with a height lower than 10 m |
| multifamily | RES-MR | (Semi-) Detached residential structure and attached small-scale residential or (rarely) commercial structure with a height between 10 and 30 m |
| commercial_innercity | RCMU | Detached and attached medium to large residential, commercial or office structure lower than 30 m |
| highrise | RCMU-HR | Detached and attached medium to large residential, commercial or office structure between 30 and 75 m |
| skyscraper | RCMU-SKY | Detached and attached medium to large residential, commercial or office structure higher than 75 m (includes skyscrapers) |
| commercial_industrial | C/I | Attached or detached light small-scale commercial, industrial or office structure and large industrial or retail and heavy industry |
| lightweight | MLB | Mobile homes and detached light-weight buildings such as wooden cabins, huts, garages, etc. |

For each building type, Baumgart et al. Provide material intensities for CONUS. For the single family category, these material intensities are reported separately for different climate zones covering the CONUS (Table A.1.2). We used climate zones from the U.S. Department of Energy (https://basc.pnnl.gov/building-assemblies) to apply the climate-zone specific material intensities.

**Appendix 2. Modelling combustible mass of building contents**

Frishcosy et al. (2021) propose a method to estimate the combustible mass of the built environment based on a relatively simple set of input parameters. These parameters include the size (i.e., the areal extent) and the function (i.e., the land use) of a specific unit of land, which could be a census unit or a cadastral parcel of homogeneous land use. For a given land use type, Frishcosy et al. (2021) provide fuel load densities (**FLD**), estimating the energy density of fuel materials in MJ/m². Thus, for a given areal unit $a$ [m²] of known land use $lu_a$, the fuel energy (**FE**) measured in MJ can be estimated as

$$FE_a = a \cdot FLD_{lu}$$

We consider the following land use classes and their associated fuel load densities, as provided by Frishcosy et al., representing median values from various scientific studies.

*Table A2.1. Fuel load densities per land use category. Source: Frishcosy et al. (2021).*

| Land use type | Fuel load density (MJ/m2) |
|---|---|
| Agriculture | 450 |
| Business | 500 |
| Commercial | 1400 |
| Educational | 734 |
| Entertainment | 393 |
| Industrial | 1018 |
| Residential | 848 |
| Utilities | 157 |

Subsequently, based on the fuel energy $FE_a$, we can estimate the mass of the fuels i.e., the combustible mass (*CM*) within the areal unit *a*. To do so, we make two assumptions:

1) Each land use category *lu* is associated with a specific composition of combustible materials. Thus, for each combustible material *m*, we can estimate its energy contribution $EC_m$, measured in %, as specified in Table A2.2.
2) For each material *m*, we can estimate its calorific value $CV_m$, i.e., its fuel energy per mass unit, measured in MJ/kg. These calorific values are shown in Table A.2.3.

*Table A2.2. Relative fuel energy contributions (EC) of various materials (m) per land use category. Source: Frishcosy et al. (2021).*

| Land use | Wood | Paper | Plastic | Cloth |
|---|---|---|---|---|
| Agriculture | 71% | 3% | 12% | 4% |
| Business | 71% | 3% | 12% | 4% |
| Commercial | 42% | 8% | 32% | 8% |
| Educational | 42% | 8% | 32% | 8% |
| Entertainment | 42% | 8% | 32% | 8% |
| Industrial | 24% | 11% | 51% | 10% |
| Residential | 71% | 3% | 12% | 4% |
| Utility | 58% | 3% | 15% | 9% |

*Table A.2.3. Calorific values associated with specific building materials. Source: Frishcosy et al. (2021).*

| Material | Wood | Plastic | Paper | Cloth |
|---|---|---|---|---|
| CV (MJ/kg) | 15.72 | 35.27 | 15.63 | 21.15 |

Based on the fuel energy $FE_a$, and the calorific values of the materials $m$ contained in the areal unit $a$, we can then estimate the total combustible mass $CM_a$ as

$$CM_a = \sum_{m=1}^{7} EC_m \cdot \frac{FE_a}{CV_m}$$

With the total combustible mass being the proportional contributions of fuel energy $FE_a$, divided by the material-specific calorific value $CV_m$ for each of the seven material types $m$. Based on this method, Frishcosy et al. (2021) obtain the combustible mass within census units associated with a given land use type. While this is a valid approach for the estimation of urban fuels and their distribution at the level of an individual city, it has several shortcomings: (a) the approach potentially ignores the heterogeneity of land use within a given areal unit, (b) census units can be very large in rural areas, as they are based on demographic criteria (e.g., a minimum population size), and (c) the temporal inconsistency of census boundaries impedes the analysis of changes in urban fuels over time.

For the above reasons, we implemented a modified version of the method proposed by Frishcosy et al., that aims to account, in part, for these shortcomings.

***First,*** we use the individual property as our analytical unit. As the records in ZTRAX typically represent individually owned real estate objects, such as buildings, or building units (e.g., individually owned apartments within a multi-apartment building), our analytical unit can be associated with a single, homogeneous land use (functional) class. Thus, we estimate the combustible mass at the building / building unit level. **Second**, we account for the vertical component of the built environment. While the method of Frishcosy et al. (2021) is based on planar areal information, we make use of the indoor area attribute provided in ZTRAX, indicating the total area across all stories of the building. ***Third***, we aggregate the property-level combustible mass into a regular grid, consisting of grid cells of 250 m × 250 m, and thus, use an enumeration unit that is independent of an imposed zoning of land use classes or census units that are defined by different criteria. The use of grid cells as enumeration units allows for multi-temporal urban fuel modelling within spatial entities that are consistent over time, enabling both cross-sectional as well as longitudinal analyses at fine spatial scale. **Fourth,** we make use of the temporal information provided in ZTRAX, i.e., the construction year of each building. The construction year allows for selecting the buildings that existed in a given year. Thus, we use this temporal attribute to create gridded surfaces measuring the urban fuel at an arbitrary point in time.

Hence, we obtained a tuple for each of the > 130,000,000 property records $Z$ in ZTRAX:

$$Z = (X_Z, Y_Z, T_Z, A_Z, LU_Z, CM_Z, FE_Z, CM_{m=1}, \ldots, CM_{m=7})$$

consisting of the geolocation of each property ($X,Y$), its construction year $T$, the indoor area $A$, the land use class $LU$, the total combustible mass $CM$, the corresponding fuel energy $FE$, as well as the components of the combustible mass components $CM_m$ of each of the seven material types $m$ contributing to the property.

We used these tuples as input to create gridded surfaces measuring the combustible mass of the building stock for each year from 1999 to 2020 for most parts of the CONUS, and within temporally consistent grid cells of 250 m spatial resolution. The records were first stratified by their construction year $T_Z$, and the allocation of a property record to its underlying grid cell was done based on the geolocation ($X_Z Y_Z$). We then aggregated (i.e., summed up) the combustible mass $CM_Z$, for all properties within each grid cell and used this value to assign the cell value during the rasterization; this process was carried out for each point in time. Following the same process, we created gridded surfaces of the fuel energy $FE_z$ and for the fuel energy components for the different building materials under consideration.

## Appendix 3. Backcasting results over time and across the rural-urban continuum

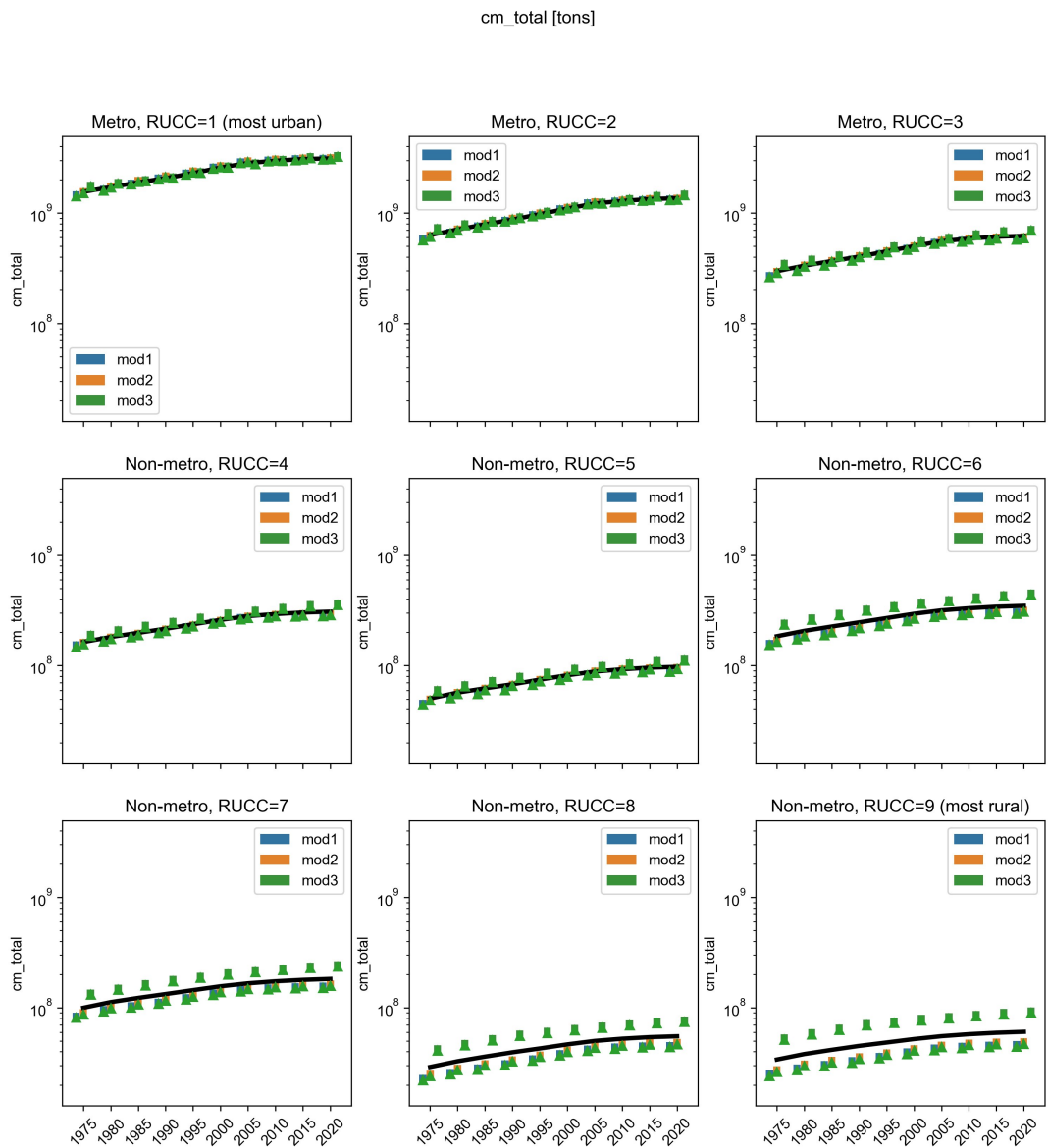

*Figure A3.1. Spread of total combustible mass across rural-urban continuum codes, for the three backcasting models.*

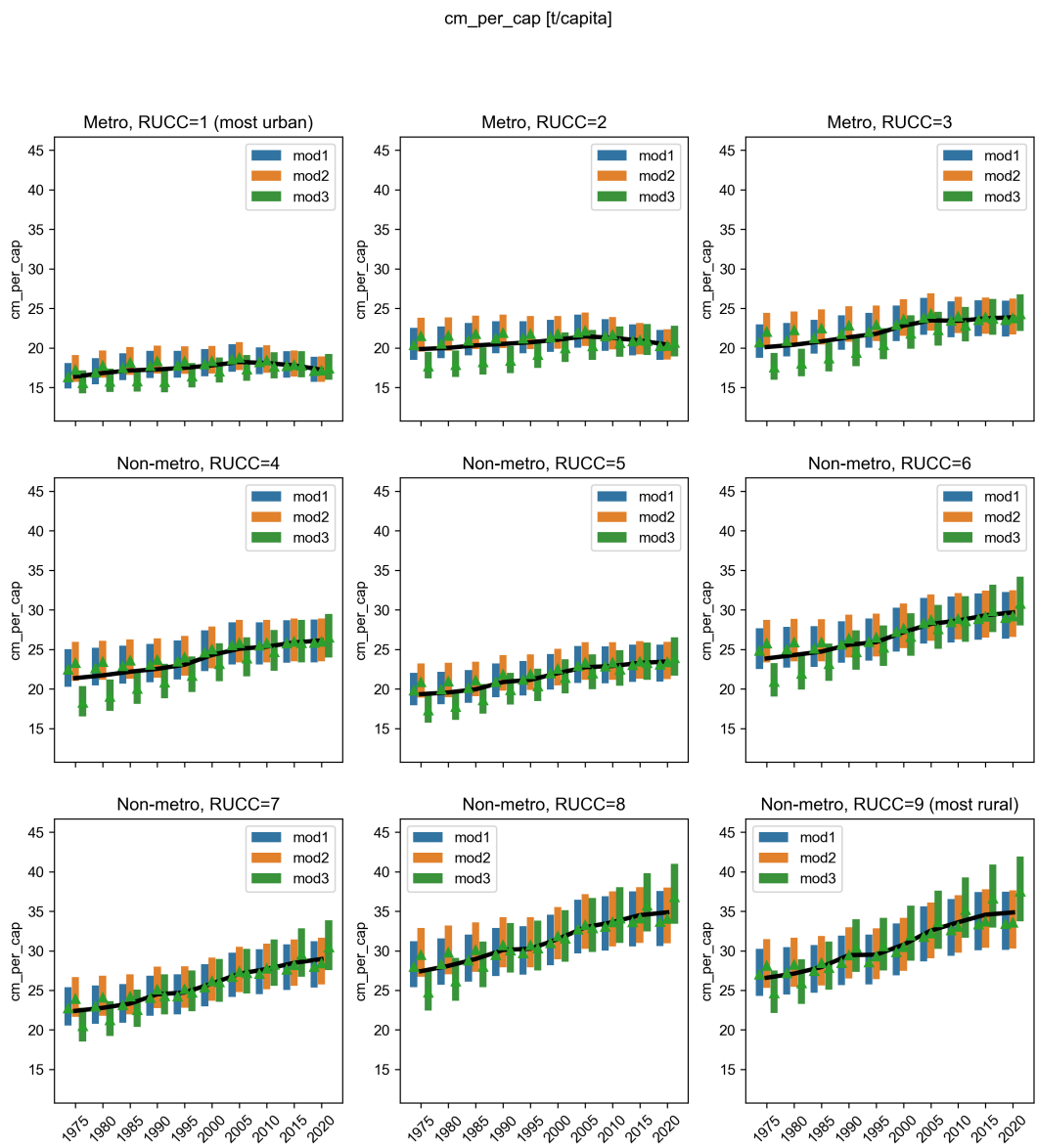

*Figure A3.2. Spread of combustible mass per capita across rural-urban continuum codes, for the three backcasting models.*

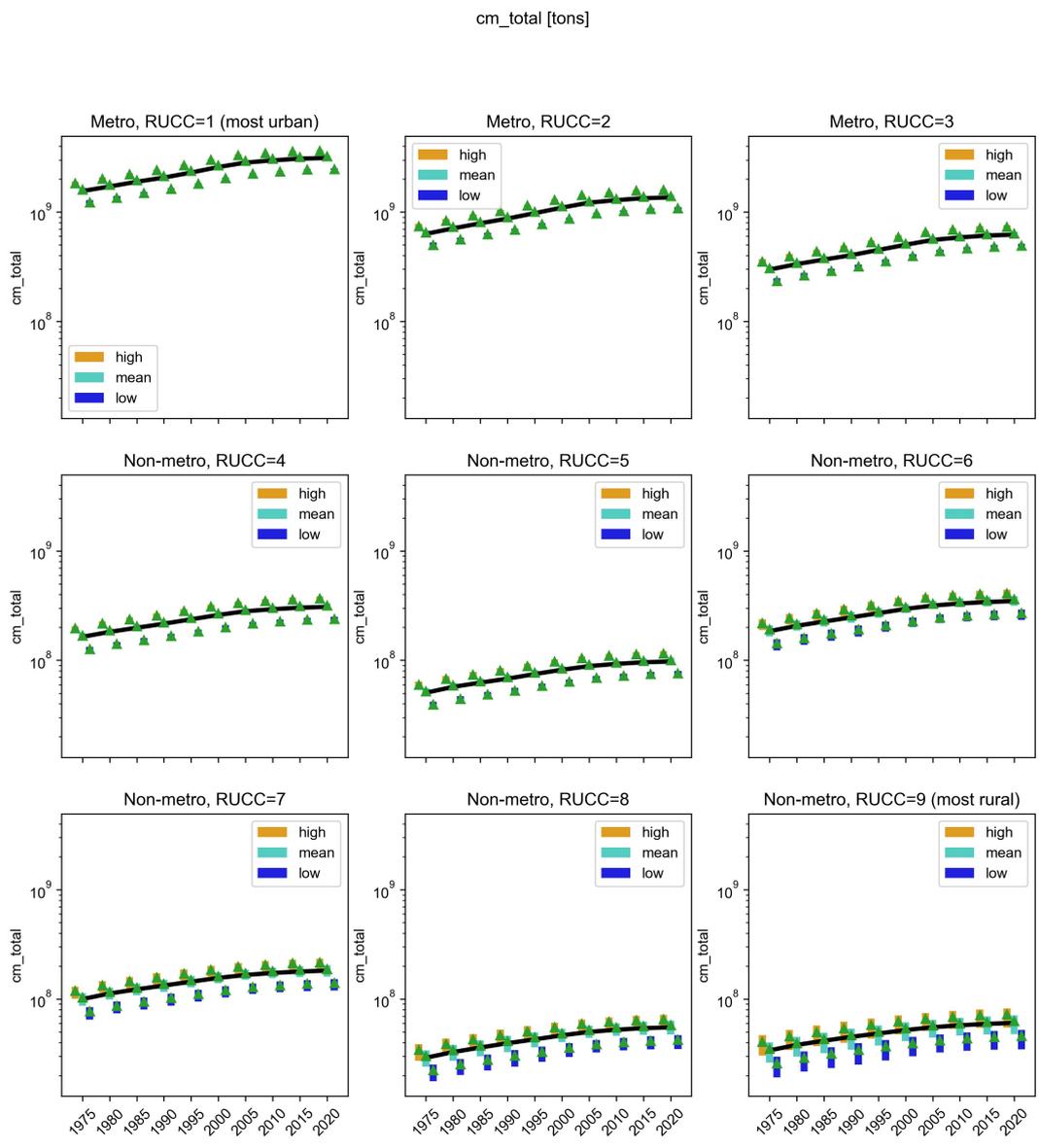

*Figure A3.3. Spread of total combustible mass across rural-urban continuum codes, for the three building mass scenarios.*

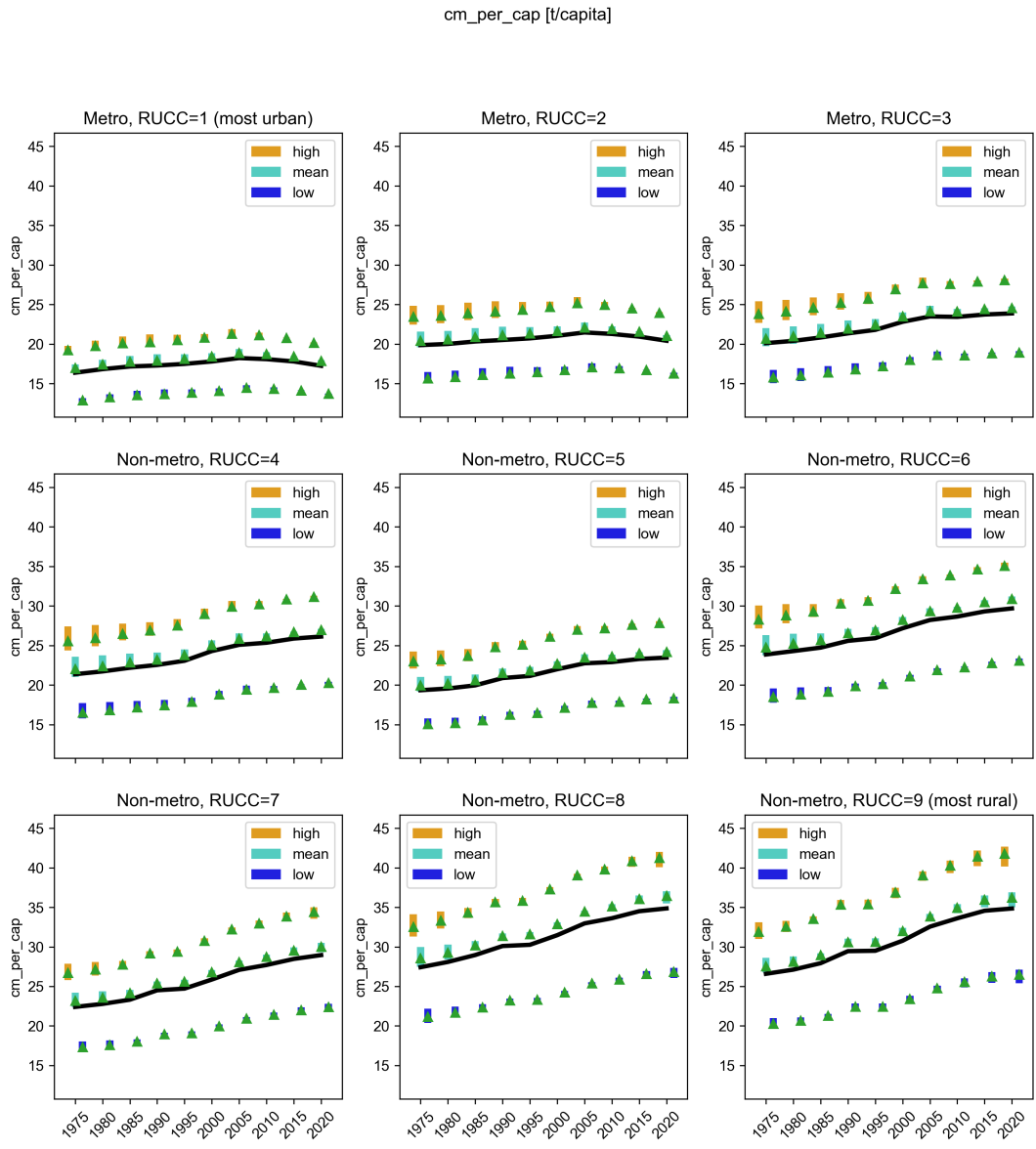

*Figure A3.4. Spread of combustible mass per capita across rural-urban continuum codes, for the three building mass scenarios.*